\documentclass[10pt,english]{article}
\usepackage[T1]{fontenc}
\usepackage[latin1]{inputenc}
\usepackage{geometry}
\usepackage{graphicx}

\makeatletter

\providecommand{\tabularnewline}{\\}

\usepackage{graphicx}

\usepackage{babel}
\makeatother
\begin{document}

\title{\textbf{Case studies of atomic properties using coupled-cluster and
unitary coupled-cluster methods}}

\author{Chiranjib Sur , Rajat K. Chaudhuri, B. P. Das\\
\emph{NAPP Group,} \emph{Indian Institute of Astrophysics, Bangalore
560 034, India}\\
D. Mukherjee\\
\emph{Indian Association for the Cultivation of Science, Kolkata -
700 032, India}}

\maketitle
\begin{abstract}
The magnetic dipole and electric quadrupole hyperfine constants of
Aluminium ($^{27}Al$) atom are computed using the relativistic coupled
cluster (CC)  and unitary coupled cluster (UCC) methods. Effects of
electron correlations are investigated using different levels of CC
approximations and truncation schemes. The ionization potentials,
excitation energies, transition probabilities, oscillator strengths
and nuclear quadrupole moment are computed to assess the accuracy
of these schemes. The nuclear quadrupole moment obtained from the
present CC and UCC calculations in the singles and doubles approximations
are 142.5 mbarn and 141.5 mbarn respectively. The discrepancies between
our calculated IPs and EEs and their measured values are better than
0.3\%. The other one-electron properties reported here are also in
excellent agreement with the measurements. 

\textbf{PACS number(s)} : 31.15.Ar, 31.15.Dv, 31.25.Jf, 32.10.Fn
\end{abstract}

\section{\label{intro}Introduction}

Theoretical studies of properties like hyperfine coupling constants
and transition probabilities are stringent tests of the accuracies
of atomic wave functions. The former is sensitive to the nuclear region
while the latter crucially depends on the wavefunctions at large distances.
High precision calculations of these properties require a rigorous
incorporation of correlation effects \cite{rajat-jcp} and in some
cases even relativistic effects. In particular, the hyperfine coupling
constants and transition electric dipole ($E1$) moments calculations
are relevant to the studies of parity non-conservation (PNC) in atoms
as PNC transition amplitudes involve both short range electro-weak
interaction and $E1$ transition moments \cite{pnc-phys-rep}. 

The relativistic and electron correlation effects can be incorporated
in many-electron systems through a variety of many-body methods. Among
these approaches, the relativistic coupled cluster (RCC) method has
emerged as one of the most powerful and effective tool for a high
precision description of electron correlations in many-electron systems
\cite{kaldor-book}. Coupled-cluster (CC) is an all-order non-perturbative
theory, and therefore, the higher order electron correlation effects
can be incorporated more efficiently than using the order-by-order
diagrammatic many-body perturbation theory (MBPT) \cite{lindgren-book}.
The CC method is size-extensive \cite{size-extensive}, a property
which has been found to be crucial for an accurate determination of
state energies, bond cleavage energies for molecules and related spectroscopic
constants. Since the order-by-order MBPT expansion terms are directly
related to the terms in the CC wavefunction (as the latter is an all-order
version of the former scheme), the CC results can be improved by adding
certain important omitted diagrams by computing the corresponding
low order MBPT diagrams to all order. 

In this paper, we report our calculations of the magnetic dipole and
electric quadrupole hyperfine constants ($A$ and $B$ respectively)
for the lowest two $^{2}P_{3/2}$ states ($3^{2}P_{3/2}$ and $4^{2}P_{3/2}$)
of $^{27}Al$ obtained using the RCC method. We also present ionization
potentials (IPs), transition energies (EEs), transition probabilities
and oscillator strengths of $^{27}Al$. Effects of electron correlations
on these quantities are investigated using different levels of CC
approximation. We compare atomic properties of $^{27}Al$ obtained
from CC and UCC methods to assess the relative performance and accuracy
of these two schemes. The UCC and its variants \cite{kutzelnigg,spal-83,spal-84,watts}
were developed almost two decades ago to incorporate higher order
electron correlation effect systematically. Recently we have applied
the relativistic UCC to atomic systems for the first time to calculate
properties like lifetime of excited states \cite{csur-ucc}. To our
knowledge, no prior UCC calculations are available for $^{27}Al$.

The nuclear quadrupole moment ($Q$) of $^{27}Al$ is of interest
in several research areas \cite{pyykko-rev}. The electric quadrupole
hyperfine constant ($B$) of $^{27}Al$ was measured in ionic crystals
\cite{pound,liao} and in metallic alloys \cite{fukai} and the value
of $Q$ is extracted by combining the calculated electric field gradient
($q$) at the nucleus with the measured value of $B$. $Q$ is also
obtained from studying $AlF$ and $AlCl$ molecules \cite{mol-exp}.
Pernpointner and Visscher \cite{pernpointner} have obtained the value
of $Q$ for $Al$, by studying $AlF$, $AlCl$ and $AlBr$ molecules
using fully relativistic CCSD(T) theory. The value of $Q$ is also
obtained from the muonic x-ray \cite{muon-exp1,muon-exp2} and nuclear
scattering experiments \cite{nucl-scatt}.

In 1976, Rogers \emph{et al} \cite{rogers} employed the second order
MBPT method to determine the nuclear quadrupole moment $Q$ of $^{27}Al$.
Later, Sundholm and Olsen \cite{sundholm-prl} calculated $Q$ for
the $^{2}P_{3/2}$ state of $Al$ using the multi-configuration Hartree-Fock
(MCHF) approach \cite{sundholm-cpl}. Nuclear structure calculations
of $Q$ have also been carried out \cite{nucl-th-1,nucl-th-2,nucl-th-3,nucl-th-4}.
The discrepancies between the calculated and measured values of $Q$
suggest that inclusion of higher order electron correlation effects
is necessary to improve the existing calculations. Our present work
is motivated by this consideration. In this work, we have compared
our calculated $Q$ value of the with all the available calculated
and measured values.

Section \ref{methods} briefly reviews the CC method. Computational
details and results are discussed in the subsequent sections \ref{comp}
and \ref{results} respectively. Finally in the last section we highlight
the findings of our work.

\section{\label{methods}Methodology}

Since the coupled cluster methods used in this work are discussed
elsewhere \cite{kutzelnigg,spal-83,spal-84,watts,lindgren,dm-spal}
in details, we only outline the essential features of the method here.

\noindent In this work, we employ the straight forward extension of
non-relativistic coupled cluster theory to the relativistic regime
by adopting the no-virtual-pair approximation (NVPA) along with appropriate
modification of orbital form and potential terms \cite{eliav}. We
begin with Dirac-Coulomb Hamiltonian ($H$) which is expressed as

\begin{equation}
H=\sum_{i=1}^{N}\left[c\vec{\alpha_{i}}\cdot\vec{p}_{i}+\beta mc^{2}+V_{N}(r_{i})\right]+\sum_{i<j}^{N}\frac{e^{2}}{r_{ij}}\,.\label{dc}\end{equation}
The normal order form of the above Hamiltonian is given by

\noindent \begin{equation}
H=H_{N}-\langle0|H|0\rangle=\sum_{ij}\langle i|f|j\rangle\left\{ a_{i}^{\dagger}a_{j}\right\} +\frac{1}{4}\sum_{i,j,k,l}\langle ij||kl\rangle\left\{ a_{i}^{\dagger}a_{j}^{\dagger}a_{l}a_{k}\right\} ,\label{eq1}\end{equation}
 where \begin{equation}
\langle ij||kl\rangle=\langle ij|\frac{1}{r_{12}}|kl\rangle-\langle ij|\frac{1}{r_{12}}|lk\rangle.\label{eq2}\end{equation}
 The valence universal Fock space open-shell coupled cluster method
is employed which begins with the decomposition of the full many-electron
Hilbert space of dimension $N$ into into a reference space $\mathcal{M}_{0}$
of dimension $M\ll N$, defined by the projector $P$, and its orthogonal
complement $\mathcal{M}_{0}^{\perp}$ associated with the projector
$Q=1-P$. A valence universal wave operator $\Omega$ is then introduced
which satisfies \begin{equation}
|\Psi_{i}\rangle=\Omega|\Psi_{i}^{(0)}\rangle,\;\;\; i=1,\ldots,M\label{eq3}\end{equation}
 where $|\Psi_{i}^{(0)}\rangle$ and $|\Psi_{i}\rangle$ are the \emph{unperturbed}
and \emph{the exact} wave functions of the $i$th eigenstate of the
Hamiltonian, respectively. The wave operator $\Omega$, which formally
represents the mapping of the reference space $\mathcal{M}_{0}$ onto
the target space $\mathcal{M}$ spanned by the $M$ eigenstates $|\Psi_{i}\rangle$,
has the properties \begin{equation}
\Omega P=\Omega,\;\; P\Omega=P,\;\;\Omega^{2}=\Omega.\label{eq4}\end{equation}

\noindent With the aid of the wave operator $\Omega$, the Schr\"{o}dinger
equation for the $M$ eigenstates of the Hamiltonian correlating with
the $M$-dimensional reference space, i.e., \begin{equation}
H|\Psi_{i}\rangle=E_{i}|\Psi_{i}\rangle,\;\;\; i=1,\ldots,M,\label{eq5}\end{equation}
 is transformed into a generalized Bloch equation, \begin{equation}
H\Omega P=\Omega H\Omega P=\Omega PH_{\mathrm{eff}}P,\label{eq6}\end{equation}
 where $H_{\mathrm{eff}}\equiv PH\Omega P$ is the effective Hamiltonian.
Once Eq. (\ref{eq6}) is solved for the wave operator $\Omega$, the
energies $E_{i}$, $i=1,\ldots,M$, are computed by diagonalizing
the effective Hamiltonian $H_{\mathrm{eff}}$ in the $M$-dimensional
reference space $\mathcal{M}_{0}$. Following Lindgren's formulation
of open-shell CC \cite{lindgren}, we express the valence universal
wave operator $\Omega$ as \begin{equation}
\Omega=\{\exp(\sigma)\},\label{eq7}\end{equation}
 and $\sigma$ being the excitation operator and curly brackets denote
the normal ordering. 

The operator $\sigma$ has two parts, one corresponds to the core
sector and the other to the valence sector. In the coupled-cluster
singles and double (CCSD) excitation approximation the excitation
operator for the core sector is given by

\begin{equation}
T=T_{1}+T_{2}=\sum_{ap}\left\{ a_{p}^{\dagger}a_{a}\right\} t_{a}^{p}+\frac{1}{2}\sum_{abpq}\left\{ a_{p}^{\dagger}a_{q}^{\dagger}a_{b}a_{a}\right\} t_{ab}^{pq}\,,\label{core-T}\end{equation}
$t_{a}^{p}$ and $t_{ab}^{pq}$ being the amplitude corresponding
to single and double excitations respectively. In UCC theory the core
excitation operator has a unitary form and is represented as $T-T^{\dagger}$.
For a single valence system the excitation operator the valance sector
turns out to be $\exp(S)=\left\{ 1+S\right\} $and

\begin{equation}
S_{k}=S_{1k}+S_{2k}=\sum_{k\neq p}\left\{ a_{p}^{\dagger}a_{k}\right\} s_{k}^{p}+\sum_{bpq}\left\{ a_{p}^{\dagger}a_{q}^{\dagger}a_{b}a_{k}\right\} s_{kb}^{pq}\,,\label{open-S}\end{equation}
where $s_{k}^{p}$ and $s_{kb}^{pq}$ denotes the single and double
excitation amplitudes for the valance sectors respectively. In Eqs.
(\ref{core-T}) and (\ref{open-S}) we denote the core (virtual )
orbitals by $a,b,c...\,(p,q,r...)$ respectively and $v$ corresponds
to the valance orbital. In the unitary counterpart of CCSD, \emph{i.e.}
in UCCSD, since the core excitation operator also contains a de-excitation
part (denoted by $T^{\dagger}$) it can be shown that for a given
approximation the UCC theory contains certain higher excitations effects
which is not present in the CC theory \cite{csur-ucc}.

\subsection{\label{1-eprop}Computation of one-electron properties}

\noindent We now present the method for computing the matrix-element
of sum of one-body operator $O=\sum_{i=1}^{N}o_{i}$ that utilizes
the structure $\Omega=\{\exp(\sigma)\}$. In this approach, the CC-equations
are first solved to determine the $\sigma$ cluster amplitudes and
then the matrix-element of a one-body operator is computed through
the following relation: 

\begin{equation}
O_{fi}=\frac{\langle\Psi_{f}|O|\Psi_{i}\rangle}{\sqrt{\langle\Psi_{f}|\Psi_{f}\rangle}\sqrt{\langle\Psi_{i}|\Psi_{i}\rangle}},\label{eq8}\end{equation}

\noindent where $|\Psi_{k}\rangle$ denotes the exact $k$-th state
wave-functions. It can be shown that the substitution of the expression
for the exact wave-functions $|\Psi_{i}\rangle$ and $|\Psi_{f}\rangle$
in Eq.(\ref{eq8}) explicitly cancels out spurious disconnected terms
from the above expression which reduces to 

\begin{equation}
O_{fi}=\frac{\langle\Psi_{f}|O|\Psi_{i}\rangle_{c}}{\sqrt{\langle\Psi_{f}|\Psi_{f}\rangle_{c}}\sqrt{\langle\Psi_{i}|\Psi_{i}\rangle_{c}}},\label{eq9}\end{equation}
where subscript $c$ refers to the `connected' terms.

\subsection{Magnetic dipole and electric quadrupole hyperfine constants}

The interaction between the various moments of the nucleus and the
electrons of an atom are collectively referred to as hyperfine interactions
\cite{lindgren-book}. Here we will briefly present and outline of
the the magnetic dipole ($A$), electric quadrupole ($B$) hyperfine
constants and the nuclear quadrupole moment ($Q$).

For a state $\left|IJFM_{F}\right\rangle $the magnetic dipole hyperfine
constant $A$ is defined as 

\begin{equation}
A=\mu_{N}\left(\frac{\mu_{I}}{I}\right)\frac{\left\langle J\right\Vert T^{(1)}\left\Vert J\right\rangle }{\sqrt{J(J+1)(2J+1)}},\label{mag-dip}\end{equation}
where $\mu_{I}$ is the nuclear dipole moment defined in units of
Bohr magneton $\mu_{N}$; $\mathbf{I}$ and $\mathbf{J}$ are the
total angular angular momentum for the nucleus and the electron state
respectively and  $\mathbf{F}=\mathbf{I}+\mathbf{J}$ with the projection
$M_{F}$. The electric quadrupole hyperfine constant $B$ for the
same state is defined as 

\begin{equation}
B=2eQ\left[\frac{2J(2J-1)}{(2J+1)(2J+2)(2J+3)}\right]^{1/2}\left\langle J\right\Vert T^{(2)}\left\Vert J\right\rangle ,\label{el-quad}\end{equation}
where $Q$ denotes the nuclear quadrupole moment.

\noindent The single particle forms ($t^{(k)}$) of the operator $T^{(k)}$($k=1,2$)
are taken from Cheng's paper \cite{cheng} and are represented as

\begin{equation}
T_{q}^{(1)}=\sum_{q}t_{q}^{(1)}=\sum_{j}-ie\sqrt{\frac{8\pi}{3}}r_{j}^{-2}\overrightarrow{\alpha_{j}}\cdot\mathbf{Y}_{1q}^{(0)}(\widehat{r_{j}})\label{sing-t1}\end{equation}
and

\begin{equation}
T_{q}^{(2)}=\sum_{q}t_{q}^{(2)}=\sum_{j}-er_{j}^{-3}C_{q}^{(2)}(\widehat{r_{j}}).\label{sing-t2}\end{equation}
Here $\overrightarrow{\alpha}$ is the Dirac matrix and $\mathbf{Y}_{kq}^{\lambda}$
is the vector spherical harmonics and $C_{q}^{(k)}=\sqrt{\frac{4\pi}{(2k+1)}}Y_{kq}$.
In Eq.(\ref{sing-t1}) the index $j$ refers to the $j$-th electron
of the atom and $e$ is the magnitude of the electronic charge.

\subsection{\label{tran-prob}Electric dipole transition probabilities and oscillator
strengths}

The transition probability $A_{f\leftarrow i}$ (in sec$^{-1}$) and
oscillator strength $f_{if}$ (in a.u.) for the electric dipole allowed
transitions are given by \cite{sobelman}

\begin{equation}
A_{f\leftarrow i}=\frac{2.0261\times10^{18}}{g_{f}\lambda^{3}}S_{f\leftarrow i}\label{Afi}\end{equation}
 and 

\begin{equation}
f_{if}=1.499\times10^{-16}\frac{g_{f}}{g_{i}}\lambda^{2}A_{f\leftarrow i}\label{os-strength}\end{equation}
 respectively. Here, $\lambda$ is the wave length in $\textrm{Å}$
and $g_{f}(g_{i})\equiv(2J+1)$ is degeneracy of the upper (lower)
level. The quantity $S_{f\leftarrow i}$ is the $E1$ line strengths
(in atomic units), respectively. The line strengths $S_{f\leftarrow i}$
is defined as 

\begin{equation}
S_{f\leftarrow i}=D_{if}\times D_{fi}\,,\label{line-strngth}\end{equation}
 where the electric dipole $D_{fi}$ matrix elements is given by

\begin{equation}
D_{fi}=C(f,i)\int dr\left[P_{f}(r)P_{i}(r)+Q_{f}(r)Q_{i}(r)\right]r\,,\label{dipole}\end{equation}
 with 

\begin{equation}
C(f,i)=(-1)^{j_{f}+1/2}\left(\begin{array}{ccc}
j_{f} & 1 & j_{i}\\
1/2 & 0 & -1/2\end{array}\right)\sqrt{(2j_{f}+1)(2j_{i}+1)}\,.\label{c-coeff}\end{equation}

\section{\label{comp}Computational Details}

The Fock-space relativistic coupled cluster method is applied to compute
the ground and excited state energies of $Al$. The Dirac-Fock equations
are first solved for the alkali metal ion $M^{+}$, which defines
the (0-hole,0-particle) sector of the Fock space. The ion is then
correlated using the closed shell CCSD/LCCSD, after which one-electron
is added following the Fock-space scheme 

\[
M^{+}(0,0)\longrightarrow M^{+}(0,1).\]
 Here LCCSD corresponds to linearized coupled-cluster in singles and
doubles. Both the DF and relativistic CC programs utilize the angular
momentum decomposition of the wave-functions and CC equations. Using
the Jucys- Levinson-Vanagas (JLV) theorem \cite{jlv}, the Goldstone
diagrams are expressed as a products of angular momentum diagrams
and reduced matrix element. This procedure simplifies the computational
complexity of the DF and relativistic CC equations. We use the kinetic
balance condition to avoid the {}``variational collapse'' \cite{kinetic balance-lee}.

In the actual computation, the DF ground state and excited state properties
of $Al$ are computed using the finite basis set expansion method
(FBSE) \cite{rajat-gauss} with a large basis set of $(32s28p24d15f)$
Gaussian functions of the form \begin{equation}
F_{i,k}(r)=r^{k}\cdot e^{-\alpha_{i}r^{2}}\label{eq20}\end{equation}
 with $k=0,1,\dots$ for $s,p,\dots$ type functions, respectively.
For the exponents, the even tempering condition \begin{equation}
\alpha_{i}=\alpha_{0}\beta^{i-1}\label{eq21}\end{equation}
 is applied. Here, $N$ is the number of basis functions for a specific
symmetry. The self-consistent DF orbitals are stored on a grid. It
is assumed that virtual orbitals with high energies do not contribute
significantly to properties like IPs. In the CCSD calculations, we
therefore truncate the virtual $s$, $p$, $d$ and $f$ orbitals
above 1000 a.u., 1000 a.u., 500 a.u. and 100 a.u., respectively. Single
and double excitations from all the core orbitals to valence or virtual
orbitals are considered.

\begin{table}[ht]

\caption{\label{tran-en}Transition energies (in cm$^{-1}$) of $Al$ atom.
IP is the ionization potential, EE denotes the excitation energies
with respect to the $^{2}P_{1/2}$ ground state.}

\begin{center}\begin{tabular}{lcccccr}
\hline 
&
 Dominant &
 State &
 LCCSD &
 CCSD &
 UCCSD &
 Expt\cite{moore}\tabularnewline
&
 Configuration &
&
&
&
&
\tabularnewline
\hline
\hline 
IP &
{[}Mg{]}$3p_{1/2}$&
 $^{2}P_{1/2}$&
 48194.92 &
 48155.42 &
 48211.83 &
 48279.16 \tabularnewline
&
&
&
&
&
&
\tabularnewline
 EE &
{[}Mg{]}$3p_{3/2}$&
 $^{2}P_{3/2}$&
 133.94 &
 114.75 &
 114.55 &
 112.04 \tabularnewline
&
{[}Mg{]}$4s_{1/2}$&
 $^{2}S_{1/2}$&
 24802.66 &
 24937.55 &
 24988.00 &
 25347.69 \tabularnewline
&
&
&
&
&
&
\tabularnewline
&
{[}Mg{]}$4p_{1/2}$&
 $^{2}P_{1/2}$&
 32464.50 &
 32521.12 &
 32572.34 &
 32949.84 \tabularnewline
&
{[}Mg{]}$4p_{3/2}$&
 $^{2}P_{3/2}$&
 32481.26 &
 32537.68 &
 32588.94 &
 32965.67 \tabularnewline
\hline
\hline 
&
&
&
&
&
&
 \tabularnewline
\end{tabular}\end{center}
\end{table}

\section{\label{results}Results and Discussions}

Table \ref{tran-en} compares the IP and EE of $Al$ computed using
different CC methods with the experiment \cite{moore}. It can be
seen from this table that UCCSD calculations of the IP and EEs are
more accurate than the CCSD and LCCSD results. Although not well understood,
the present as well as some earlier studies \cite{rajat-jcp} indicate
that the IPs computed using the LCCSD scheme are sometimes in better
agreement with the experiments than the corresponding CCSD calculations.
For instance, the $^{2}P_{1/2}(3p_{1/2})$ IP estimated using LCCSD
method is differs by $84\, cm^{-1}$ from the measured value, while
the corresponding CC value is off by $124\, cm^{-1}$. However, it
is clear from Table \ref{tran-en} that CCSD estimates the EE more
accurately than the LCCSD scheme for the low lying states.

In table \ref{qval} we present the results of our nuclear quadrupole
moment ($Q$) calculation using different CC methods with other calculations
\cite{rogers,sundholm-prl} and different measurements \cite{mol-exp,muon-exp1,muon-exp2,nucl-scatt}.
Pyykk$\mathrm{\ddot{o}}$ has reviewed, calculated and measured $Q$
values for a number of systems \cite{pyykko-rev}. Comparison of our
results with the existing data will give an indication of the potential
of the CC and UCC methods to provide accurate estimate of nuclear
properties. It is evident from table \ref{qval} that $Q$ calculated
using the second order MBPT calculations by Rogers \emph{et al.} \cite{rogers}
is far outside the experimental limits whereas the value obtained
by the restricted active space (RAS) based multi-configuration Hartree-Fock
(MCHF) \cite{sundholm-cpl} calculation is closer to the experimental
limits. The uncertainty in molecular experiment is less compared to
the muonic experiments. Although our CC and UCC results are slightly
outside the experimental limits, they could be of some importance
in determining the accurate value of $Q$ from a wide range of values.

\begin{table}[ht]

\caption{\label{qval}Comparison of the nuclear quadrupole moments $Q$ (in
mbarn) of $^{27}$Al estimated using various CC approach with the
experiment and with other theoretical calculations.}

\begin{center}\begin{tabular}{ll}
\hline 
Method &
$Q$ \tabularnewline
\hline
\hline 
LCCSD &
 146.7 \tabularnewline
CCSD &
 142.5 \tabularnewline
UCCSD &
 141.5 \tabularnewline
MBPT(2)\cite{rogers}&
 165(2) \tabularnewline
MCHF\cite{sundholm-cpl}&
 140.3(1.0) \tabularnewline
Molecular Exp\cite{mol-exp} &
146.6(1.0)\tabularnewline
Molecular Theory \cite{pernpointner}&
146.0(4)\tabularnewline
Muonic Exp.\cite{muon-exp1,muon-exp2}&
150(6)\tabularnewline
Nuclear Scattering\cite{nucl-scatt}&
155(3)\tabularnewline
Nuclear Theory\cite{nucl-th-1,nucl-th-2}&
134\tabularnewline
Nuclear Theory\cite{nucl-th-2,nucl-th-3}&
150.8\tabularnewline
Nuclear Theory\cite{nucl-th-3,nucl-th-4}&
138.9\tabularnewline
\hline
\hline 
&
\tabularnewline
\end{tabular}\end{center}
\end{table}

The MBPT(2) and MCHF results clearly indicate that the contributions
from non linear terms present in CC and UCC theories are non-negligible
and this is further supported by the results of our different CC calculations
of $Q$. The extremely accurate estimate of $Q$ offered by LCCSD
scheme is perhaps a bit fortuitous. Nevertheless, the performance
of CC, especially the UCC, outshines the MBPT(2) and MCHF treatments.
Note that the effects of partial triple and quadrupole excitations
are present in our UCC calculations. The CC and UCC theories unlike
the MCHF method are size-consistent and incorporate certain higher
order excitations that the MCHF does not at the same level of approximation.
For example, at the level of single and double (SD) excitations, the
CC theory includes not only the effect of $T_{2}$ but also $T_{2}^{2}$;
whereas the effect of $T_{2}^{2}$ can be obtained in MCHF only if
one considers the quadrupole excitations. Also, that calculation is
non-relativistic with a relativistic correction added while our calculation
is fully relativistic.

\begin{table}[ht]

\caption{\label{hyp}Magnetic dipole hyperfine (A) matrix elements ( in MHz)
of Al atom.}

\begin{center}\begin{tabular}{llllll}
\hline 
Method &
 A($3p_{1/2}$) &
 A($3p_{3/2}$) &
 A($4s_{1/2}$) &
 A($4p_{1/2}$) &
 A($4p_{3/2}$) \tabularnewline
\hline
\hline 
LCCSD &
 493.30 &
 108.39 &
 414.14 &
 55.97 &
 26.26 \tabularnewline
CCSD &
 498.06 &
 101.49 &
 405.94 &
 58.32 &
 23.09 \tabularnewline
UCCSD &
 498.33 &
 100.98 &
 407.18 &
 58.28 &
 23.12 \tabularnewline
Expt \cite{A-exp}&
502.0346(5)&
94.27723(10)&
&
&
\tabularnewline
\hline
\hline 
&
&
&
&
&
\tabularnewline
\end{tabular}\end{center}
\end{table}

The values of $A$ for the ground and excited states of $Al$ computed
using the LCCSD, CCSD and UCCSD methods are displayed in Table \ref{hyp}.
Our calculated values of $A$ agrees well with the experimental values
\cite{A-exp} for the $^{2}P_{1/2}(3p_{1/2})$ and $^{2}P_{3/2}(3p_{3/2})$
state. We also present the values of $A$ for some other low lying
states which could be useful for experimentalists. We have also computed
the electric quadrupole hyperfine constant ($B$) for two low lying
$^{2}P_{3/2}(3p_{3/2})$ and $(4p_{3/2})$ states using CCSD(UCCSD)
theory which are respectively 19.49 MHz(19.59 MHz) and 2.85 MHz(2.86
MHz) whereas the experimental value of $B$ for the $^{2}P_{3/2}(3p_{3/2})$
state is 18.915 MHz \cite{martin-B-exp,harvey-B-exp}.

\begin{figure}[ht]
\begin{center}\includegraphics{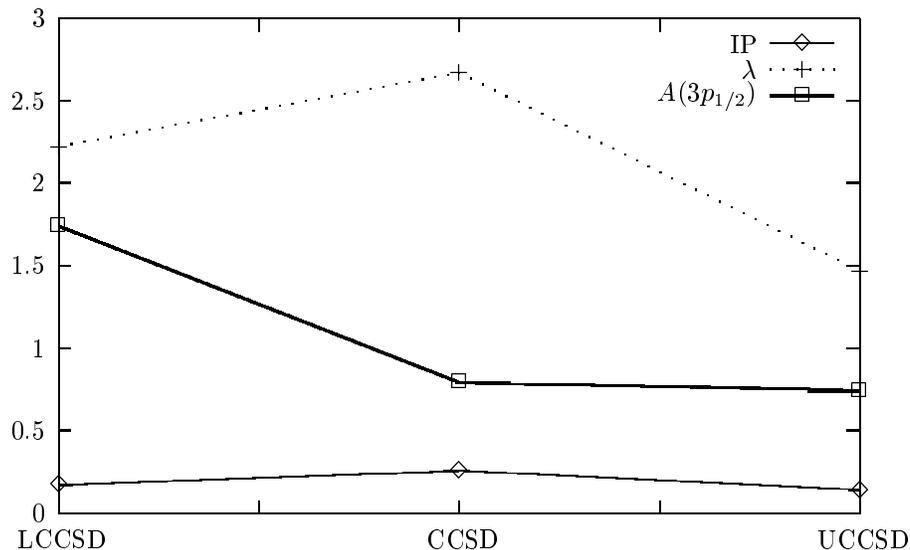}\end{center}

\caption{\label{error}Absolute error (in \%) of the computed IP, $^{2}P_{1/2}(3p_{1/2})\rightarrow^{2}S_{1/2}(4s)$
wave lengths ($\lambda$), magnetic dipole hyperfine constant ($A$)
for the $3p^{2}P_{1/2}$ state using different CC methods. }
\end{figure}

In Table \ref{tran-prop}, we compare the $3p_{1/2}\rightarrow4s$
and $3p_{3/2}\rightarrow3s$ wave lengths ($\lambda$), oscillator
strengths ($f_{ik}$), line strengths ($S_{ik}$) and transition probabilities
($A_{ik}$) obtained from LCCSD, CCSD and UCC schemes with the experiment.
Table \ref{tran-prop} shows that our computed quantities ($\lambda$,
$S_{ik}$, and $A_{ik}$) are in excellent agreement with experiment
especially those predicted by the UCC scheme. That this scheme provides
more accurate estimates of IP, EE etc. and is also evident from Figure
\ref{error}, where the absolute errors (in \%) in the computed properties
are plotted against different CC schemes.

\begin{table}[ht]

\caption{\label{tran-prop}Wave lengths $\lambda$ (in $\textrm{Å}$), line
strengths $S_{ik}\equiv|r|^{2}/a_{0}^{2}$ (in a.u.),, transition
probabilities $A_{ik}$ (in $10^{8}s^{-1}$), and oscillator strengths
$f_{ik}$ (in a.u.) for $[\mbox{Mg}]3p\rightarrow[\mbox{Mg}]4s$ transitions
of $Al$ atom.}

\begin{center}\begin{tabular}{lcccccccr}
\hline 
Method &
\multicolumn{4}{c}{{[}Mg{]}$3p_{1/2}\rightarrow4s$}&
\multicolumn{4}{c}{{[}Mg{]}$3p_{3/2}\rightarrow4s$}\tabularnewline
&
 $\lambda$&
 $S_{ik}$&
 $A_{ik}$&
 $f_{ik}$&
 $\lambda$&
 $S_{ik}$&
 $A_{ik}$&
 $f_{ik}$\tabularnewline
\hline
\hline 
LCCSD &
 4031.82 &
 3.379 &
 0.522 &
 0.127 &
 4053.72 &
 6.763 &
 1.028 &
 0.127 \tabularnewline
CCSD &
 4010.05 &
 3.292 &
 0.517 &
 0.125 &
 4028.59 &
 6.634 &
 1.028 &
 0.125 \tabularnewline
UCCSD &
 4001.92 &
 3.275 &
 0.517 &
 0.114 &
 4020.35 &
 6.600 &
 1.029 &
 0.125 \tabularnewline
 Expt\cite{tran-prob}&
 3944.01 &
 2.99 &
 0.493&
 0.115 &
 3961.52 &
 6.0 &
 0.98 &
 0.115  \tabularnewline
\hline
\hline 
&
&
&
&
&
&
&
&
\tabularnewline
\end{tabular}\end{center}
\end{table}

\section{\label{concln}Conclusion}

The relativistic open-shell coupled cluster scheme for direct energy
difference calculations and several one electron properties is presented
and applied to $Al$. In this work, we investigate the effects of
electron correlations on the ground and excited state properties using
different levels of CC approximations. We have shown that unitary
coupled cluster (UCC) method is capable of providing accurate estimates
of wave lengths, transition probabilities, oscillator strengths, nuclear
quadrupole moment, magnetic dipole and electric quadrupole hyperfine
constants for relatively light atomic systems with a single valence
electron. The calculated value of $Q$ compared to others are closer
to the experimental uncertainties than all the existing atomic and
nuclear calculations, thereby demonstrating that RCC theory of atoms
can yield accurate values of nuclear quadrupole moments. Such an inter-disciplinary
approach involving atomic and nuclear physics adds a new dimension
to this theory. 

\begin{verse}
\textbf{Acknowledgments} : One of the authors (CS) acknowledges the
BRNS for project no. 2002/37/12/BRNS.
\end{verse}

\end{document}